\documentclass[preprint,12pt]{elsarticle}

\usepackage{amssymb}
\usepackage{amsthm}
\theoremstyle{plain}
\newtheorem{thm}{Theorem}[section]

\newtheorem{prop}[thm]{Proposition}

\theoremstyle{definition}
\newtheorem{defn}{Definition}[section]

\theoremstyle{remark}

\journal{ }

\begin{document}

\begin{frontmatter}



\title{Discrete Time Elastic Vector Spaces}


\author{Pierre-Francois Marteau}

\address{VALORIA, Universit\'e de Bretagne Sud (Universit\'e Europ\'eenne de Bretagne),\\ Campus de Tohannic, 56000 Vannes, France \\ pierre-francois.marteau@univ-ubs.fr}

\begin{abstract}
This paper proposes a framework  dedicated to the construction of what we call \textit{time elastic inner products}  allowing one to embed sets
of non-uniformly sampled multivariate time series of varying lengths into vector space structures. This framework is based on a recursive definition that covers the case of multiple embedded \textit{time elastic} dimensions. We prove that such inner products exist in our framework and show how a simple instance of this inner product class operates on some toy or prospective applications, while generalizing the Euclidean inner product.

\end{abstract}


\end{frontmatter}


\section{Introduction}
\label{Intro}
Time series analysis in metric spaces has attracted much attention over numerous decades and in various domains such as biology, statistics, sociology, networking, signal processing, etc, essentially due to the ubiquitous nature of time series, whether they are symbolic or numeric. Among other characterizing tools, time warp distances (see \cite{TWED:Velichko70}, \cite{TWED:Sakoe71}, and more recently \cite{TWED:Chen04}, \cite{TWED:Marteau09} among other references) have shown some interesting robustness compared to the Euclidean  metric especially when similarity searching in time series data bases is an issue. Unfortunately, this kind of elastic distance does not enable direct construction of definite kernels which are useful when addressing regression, classification or clustering of time series. A fortiori, they do not make it possible to directly construct inner products involving some \textit{time elasticity}, which are namely able to cope with some time stretching or some time compression.  Recently, \cite{TWED:Marteau2010} have shown that it is quite easy to propose inner product with time elasticity capability at least for some restricted time series spaces, basically spaces containing uniformly sampled time series, all of which have the same lengths (in such cases, time series can be embedded easily in Euclidean spaces).\\

The aim of this paper is to derive an extension from this
 preliminary work for the construction of time elastic inner products, to achieve the construction of a time elastic inner product for a \textit{quasi-unrestricted} set
of time series, i.e. sets for which the times series are not uniformly sampled and have any lengths. Section two of the paper, following preliminary results presented in \cite{TWED:Marteau2010}, gives the main notations used throughout this paper and presents a recursive construction for inner-like products. It then gives the conditions and the proof of existence of time elastic inner products (and time elastic vector spaces) defined on a \textit{quasi-unrestricted} set of times series while explaining what we mean by \textit{quasi-unrestricted}. The third section succinctly presents some  applications, mainly to highlight some of the features of Time Elastic vector Spaces such as orthogonality.\\

\section{Discrete Time Elastic Vector Spaces}
\label{EVS}

\subsection{Sequence and sequence element}

\begin{defn}
\label{Sequence and sequence element}
Given a finite sequence $A$ we note
 $A(i)$ the $i^{th}$ element (symbol or sample) of sequence $A$. We will consider that $A(i) \in S \times T$ where $(S,\oplus_S, \otimes_S)$ is a vector space that embeds the multidimensional \textit{space} variables (e.g. $S \subset \mathbb{R}^d$, with $d \in \mathbb{N}^+$) and $T \subset \mathbb{R}$ embeds the \textit{timestamps} variable, so that we can write $A(i)=(a(i),t_{a(i)})$ where $a(i) \in S$ and $t_{a(i)} \in T$, with the condition that $t_{a(i)}> t_{a(j)}$ whenever $i>j$ (timestamps strictly increase in the sequence of samples).
$A_{i}^{j}$ with $i \leq j$ is the subsequence consisting of the $i_{th}$ through the $j_{th}$ element (inclusive) of $A$. So $A_{i}^{j}=A(i)A(i+1)...A(j)$. $\Lambda$ denotes the null element. By convention $A_{i}^{j}$ with $i>j$ is the null time series, e.g. $\Omega$.\\
\end{defn}

\subsection{Sequence set}
\begin{defn}
\label{Sequence set}
The set of all finite discrete time series is thus embedded in a spacetime characterized by a single discrete \textit{temporal} dimension, that encodes the timestamps, and any number of \textit{spatial} dimensions that encode the value of the time series at a given timestamp.
 We note $\mathbb{U}$ $= \{A_{1}^{p} | p\in \mathbb{N}\}$ the set of all finite discrete time series. $A_{1}^{p}$ is a time series with discrete index varying between $1$ and $p$. We note $\Omega$ the empty sequence (with null length) and by convention $A_{1}^{0}=\Omega$ so that $\Omega$ is a member of set $\mathbb{U}$. $|A|$ denotes the length of the sequence $A$.
Let $\mathbb{U}_p$ = $\{A \in \mathbb{U}\ | \  |A|\ \le p \}$ be the set of sequences whose length is shorter or equal to $p$. Finally let $\mathbb{U}^*$ be the set of discrete times series defined on $(S-\{0_S\}) \times T$, i.e. the set of time series that do not contain the null \textit{spatial} value. We denote by $0_S$ the null value in $S$.\\
\end{defn}

\subsection{Scalar multiplication on $\mathbb{U}^*$}
\begin{defn}
\label{Def_Otimes}
For all $A \in \mathbb{U}^*$ and all $\lambda \in \mathbb{R}$, $C = \lambda \otimes A $ $\in \mathbb{U}^*$ is such that 
for all $i \in \mathbb{N}$ such that $0\le i \le |A|$, $C(i)=(\lambda.a(i), t_{a(i)})$ and thus $|C|=|A|.$\\
\end{defn}

\subsection{addition on $\mathbb{U}^*$}
\begin{defn}
\label{Def_Oplus}
For all $(A, B) \in (\mathbb{U^*})^2$, the addition of $A$ and $B$, noted $C = A \oplus B $ $\in \mathbb{U}^*$, is defined in a constructive manner as follows:\_
Let $i,j$ and $k$ be in $\mathbb{N}$.
\begin{enumerate}[]
 \item $k=i=j=1$,
 \item As far as $1 \le i \le |A|$ and $1 \le j \le |B|$,
 \begin{enumerate}[]
 \item if $t_{a_i} < t_{b_j}$, $C(k)=(a(i),t_{a_i})$ and $i \leftarrow i+1, k \leftarrow k+1$
 \item else if $t_{a_i} > t_{b_j}$, $C(k)=(b(j),t_{b_j})$ and $j \leftarrow j+1, k \leftarrow k+1$ 
 \item else if $a_i + b_j \neq 0$, $C(k)=(a(i)+b(j),t_{a_i})$ and $i \leftarrow i+1, j \leftarrow j+1, k \leftarrow k+1$ 
 \item else $i \leftarrow i+1, j \leftarrow j+1$\\
\end{enumerate}
\end{enumerate}
\end{defn}

\begin{figure*}[!h]
\centering
\includegraphics[scale=0.75]{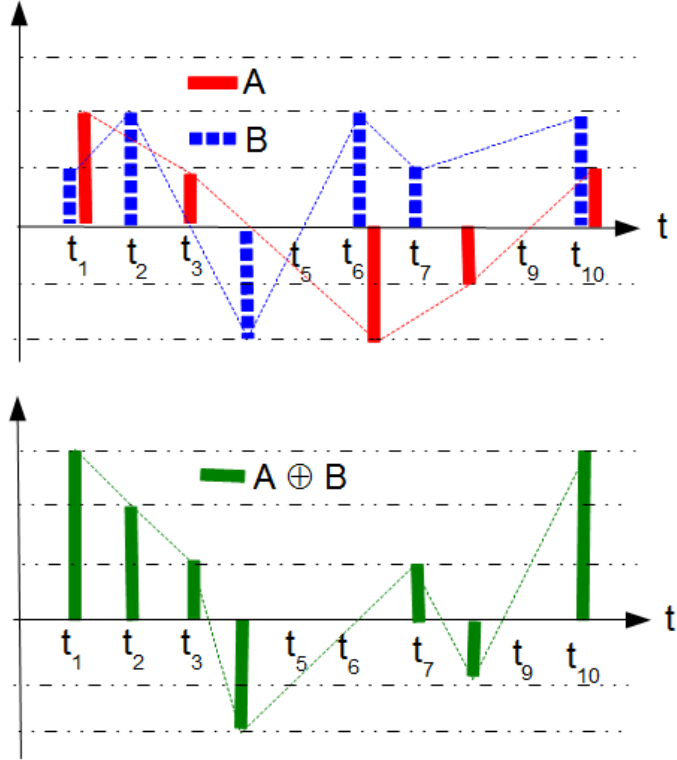}
\caption{The $\oplus$ binary operator when applied to two discrete time series of variable lengths and not uniformly sampled. Co-occurring events have been slightly separated at the top of the figure for readability purposes.}

\label{fig:AplusB}
\end{figure*}

Three comments need to be made at this level to clarify the semantic of the operator $\oplus$:
\begin{enumerate}[i)]
\item Note that the $\oplus$ addition of two time series of equal lengths and uniformly sampled coincides with the classical addition in vector spaces. Fig. \ref{fig:AplusB} gives an example of the addition of two time series that are not uniformly sampled and that have different lengths.
\item Implicitly (in light of the last case described in Def. \ref{Def_Oplus}), any sequence element of the sort $(0_S, t)$, where $0_S$ is the null value in $S$ and $t \in T$ must be assimilated to the null sequence element $\Lambda$. For instance, the addition of $A=(1,1)(1,2)$ with $B=(-1,1)(1,2)$ is $C= A\oplus B = (2,2)$: the addition of the two first sequence elements is $(0,1)$ that is assimilated to $\Lambda$ and as such suppressed in $C$.
\item The $\oplus$ operator, when restricted to the set $\mathbb{U}^*$ is reversible in that if $C= A \oplus B$ then $A = C \oplus ((-1)\otimes B)$ or $B = C \oplus ((-1) \otimes A)$. This is not the case if we consider the entire set $\mathbb{U}$.
\end{enumerate}

\subsection{Time elastic product (TEP)}
\begin{defn}
\label{tep}
A function $<.,.>: \mathbb{U}^* \times \mathbb{U}^* \rightarrow \mathbb{R}$ is called a Time Elastic Product if, for any pair of sequences $A_{1}^{p},B_{1}^{q}$, there exists a function $f: S^2 \rightarrow \mathbb{R}$, a non negative symmetric function $g: T^2 \rightarrow \mathbb{R}^+$ and three constants $\alpha$, $\beta$ and $\xi$ in $\mathbb{R}$ such that the following recursive equation holds:

\begin{eqnarray}
\label{Eq.4}
\begin{array}{ll}
 &<A_{1}^{p},B_{1}^{q}>_{tep}=  \\
 & \sum \left\{
   \begin{array}{ll}
     \alpha\cdot<A_{1}^{p-1},B_{1}^{q}>_{tep}  \\
     \beta\cdot<A_{1}^{p-1},B_{1}^{q-1}>_{tep} + f(a(p),b(q))\cdot g(t_{a(p)},t_{b(q)}) \\
     \alpha\cdot<A_{1}^{p},B_{1}^{q-1}>_{tep}  \\
   \end{array}
   \right.
  \end{array}
\end{eqnarray}
\end{defn}

This recursive definition requires defining an initialization. To that end we set, $\forall A \in \mathbb{U}^*$, $<A,\Omega>_{tep}=<\Omega,A>_{tep}=<\Omega,\Omega>_{tep}=\xi$, where $\xi$ is a real constant (typically we set $\xi=0$), and $\Omega$ is the null sequence, with the convention that ${A_i}^j=\Omega$ whenever $i>j$.  \\

It has been shown in \cite{TWED:Marteau2010} that time elastic inner products can easily be constructed from Def. \ref{tep} using the $\oplus$ and $\otimes$ operations when we restrict the set of time series to some subset containing uniformly sampled time series of equal lengths (in that case, the $\oplus$ coincides with the classical addition on $S$). For instance, definitions \ref{eqTWIP1} and \ref{eqTWIP2} recursively define two $TEP$ that are inner products on such restrictions.

\begin{defn}
\label{eqTWIP1}
\begin{eqnarray}
\begin{array}{ll}
<A_{1}^{p}, B_{1}^{q}>_{twip_1}=  \frac{1}{3}\cdot \\
\hspace*{2mm} \sum \left\{ 
		\begin{array}{ll}
     <A_{1}^{p-1},B_{1}^{q}>_{twip_1} \\
     <A_{1}^{p-1},B_{1}^{q-1}>_{twip_1} + e^{-\nu.d(t_{a(p)},t_{b(q)})}(a(p) \cdot b(q))\\
     <A_{1}^{p},B_{1}^{q-1}>_{twip_1} \\
   \end{array}
   \right.
  \end{array}
\end{eqnarray}
where $d$ is a distance, and $\nu$ a \textit{time stiffness} parameter. \\
\end{defn}

\begin{defn}
\label{eqTWIP2}
\begin{eqnarray}
\begin{array}{ll}
<A_{1}^{p}, B_{1}^{q}>_{twip_2}=  \frac{1}{1+2 \cdot e^{-\nu}}\cdot \\
\hspace*{2mm} \sum \left\{ 
		\begin{array}{ll}
     e^{-\nu}\cdot<A_{1}^{p-1},B_{1}^{q}>_{twip_2} \\
     <A_{1}^{p-1},B_{1}^{q-1}>_{twip_2} + e^{-\nu.d(t_{a(p)},t_{b(q)})}(a(p) \cdot b(q))\\
     e^{-\nu}\cdot<A_{1}^{p},B_{1}^{q-1}>_{twip_2} \\
   \end{array}
   \right.
  \end{array}
\end{eqnarray}
where $d$ is a distance, and $\nu$ a \textit{time stiffness} parameter.\\
\end{defn}

It can be shown that $<.,.>_{twip_2}$ coincides with the Euclidean inner product on the considered restrictions of $\mathbb{U}$ when $\nu \rightarrow \infty$.\\

This paper addresses the more interesting question of the existence of similar elastic inner products on the set $\mathbb{U}^*$ itself, i.e. without any restriction on the lengths of the considered time series nor the way they are sampled. If the choice of functions $f$ and $g$, although constrained, is potentially large, we show hereinafter that the choice for constants $\alpha$, $\beta$ and $\xi$ is unique. 

\subsection{Existence of $TEP$ inner products defined on $\mathbb{U}^*$}

\begin{thm}
\label{Main theorem}
$<.,.>_{tep}$ is an inner product on $(\mathbb{U}^*, \oplus, \otimes)$ iff:
\begin{enumerate}[i)]
\item $\xi = 0$.
\item $h:(S \times T) \rightarrow \mathbb{R}$ defined as $h((a,t_a))=f(a,a)\cdot g(t_a,t_a)$ is  strictly positive on $((S-\{0_S\}) \times T)$,
\item $f$ is an inner product on $(S,\oplus_S, \otimes_S)$, if we extend the domain of $f$ on $S$ while setting $f(0_S,0_S)=0$. 
\item $\alpha=1$ and $\beta=-1$,
\end{enumerate} 
\end{thm}

\subsubsection{proof of theorem \ref{Main theorem}}

\textbf{Proof of the direct implication}\\
Let us suppose first that $<.,.>_{tep}$ is an inner product defined on $\mathbb{U}^*$. Then $<.,.>_{tep}$ is positive-definite, and thus $<\Omega, \Omega>_{tep}=\xi=0$. Furthermore, for any $A=(a,t_a) \in \mathbb{U}^*$, $<A,A>_{tep}=h(a,t_a)) > 0$. Thus i) and ii) are satisfied. As $g$ is non-negative, if we set $f(0_S,0_S)=0$, $f$ is positive-definite on $S$.\\

It is also straightforward to show that $f$ is symmetric if $g$ and $<.,.>_{tep}$ are symmetric.\\

Since $\xi=0$, for any $A$, $B$, and $C \in \mathbb{U}^*$ such that $A=(a, t)$, $B(b,t)$ and $C=(c,t_c)$, we have:\\
$<A \oplus B,C>_{tep}=h((a \oplus_S b,t),(c,t_c))=f(a \oplus_S b, c).g(t,t_c)$. \\
As $<A \oplus B,C>_{tep}=<A, C>_{tep}+<B,C>_{tep}$ \\
$=h((a,t),(c,t_c))+h((b,t),(c,t_c))$\\
$=f(a,c).g(t,t_c)+f(a,c).g(t,t_c)=(f(a,c)+f(b,c)).g(t,t_c)$,\\
As $g$ is non negative, we get that $f(a \oplus_S b, c)=(f(a,c)+f(b,c))$.\\
Furthermore, $<\lambda \otimes A,C>_{tep}=h((\lambda \otimes_S a,t),(c,t_c))=f(\lambda \otimes_S a, c).g(t,t_c)$. \\
As $<\lambda \otimes A,C>_{tep}=\lambda.<A, C>_{tep}=\lambda.f(a,c).g(t,t_c)$ and $g$ is non negative, we get that $f(\lambda \otimes_S a, c)= \lambda.f(a,c)$.\\
This shows that $f$ is linear, symmetric and positive-definite. Hence it is an inner product on $(S,\oplus_S, \otimes_S)$ and iii) is satisfied.\\

Let us show that necessarily $\alpha=1$ and $\beta=-1$. To that end, let us consider any $A_{1}^p, B_{1}^q$ and $C_{1}^r$ in $\mathbb{U^*}$, such that $p>1, q>1, r>1$ and such that $t_{a_p}<t_{b_q}$, i.e.  if $X_{1}^{s}=A_{1}^{p} \oplus B_{1}^{q}$, then $X_{1}^{s-1}=A_{1}^{p} \oplus B_{1}^{q-1}$.\\
Since by hypothesis $<.,.>_{tep}$ is an inner product $(\mathbb{U}^*, \oplus, \otimes)$, it is linear and thus we can write:\\
$<A_{1}^p \oplus B_{1}^q,C_{1}^r>_{tep}=<A_{1}^p,C_{1}^r>_{tep}+<B_{1}^q,C_{1}^r>_{tep}$.\\

Decomposing $<A_{1}^p \oplus B_{1}^q,C_{1}^r>_{tep}$, we obtain:\\
$<A_{1}^p \oplus B_{1}^q,C_{1}^r>_{tep}=\alpha.<A_{1}^{p} \oplus B_{1}^{q-1}, C_{1}^r>_{tep} + $\\
$\beta.<A_{1}^{p} \oplus B_{1}^{q-1}, C_{1}^{r-1}>_{tep} + f(b_q,c_r).g(t_{b_q}, t_{c_r}) + \alpha.<A_{1}^p \oplus B_{1}^q,C_{1}^{r-1}>_{tep}$\\
As $<.,.>_{tep}$ is linear we get:\\
$<A_{1}^p \oplus B_{1}^q,C_{1}^r>_{tep}=\alpha.<A_{1}^{p},C_{1}^r>_{tep} + \alpha.<B_{1}^{q-1}, C_{1}^r>_{tep} + $\\
$\beta.<A_{1}^{p}, C_{1}^{r-1}>_{tep} + \beta.<B_{1}^{q-1}, C_{1}^{r-1}>_{tep} + f(b_q,c_r).g(t_{b_q}, t_{c_r}) + $\\
$\alpha.<A_{1}^p,C_{1}^{r-1}>_{tep} + \alpha.<B_{1}^q,C_{1}^{r-1}>_{tep}$\\
Hence,\\
$<A_{1}^p \oplus B_{1}^q,C_{1}^r>_{tep}=\alpha.<A_{1}^{p},C_{1}^r>_{tep}+\beta.<A_{1}^{p},C_{1}^{r-1}>_{tep} +$\\
$\alpha.<A_{1}^{p},C_{1}^{r-1}>_{tep} + <B_{1}^{q}, C_{1}^{r}>_{tep}$\\

If we decompose $<A_{1}^{p},C_{1}^r>_{tep}$, we get:\\
$<A_{1}^p \oplus B_{1}^q,C_{1}^r>_{tep}=(\alpha^2+\beta+\alpha)<A_{1}^p,C_{1}^{r-1}>_{tep}+\alpha.\beta.<A_{1}^{p-1},C_{1}^{r-1}>_{tep}+\alpha.f(a_p,c_r).g(t_{a_p},t_{c_r})+ \alpha^2.<A_{1}^{p-1},C_{1}^{r}>_{tep} + <B_{1}^q,C_{1}^{r}>_{tep}$\\

Thus we have to identify $<A_{1}^p,C_{1}^{r}>_{tep}=\alpha.<A_{1}^p,C_{1}^{r-1}>_{tep} + \beta.<A_{1}^{p-1},C_{1}^{r-1}>_{tep} + f(a_p,c_r).g(t_{a_p},t_{c_r}) + \alpha.<A_{1}^{p-1},C_{1}^{r}>_{tep}$\\
with $(\alpha^2+\beta+\alpha)<A_{1}^p,C_{1}^{r-1}>_{tep}+\alpha.\beta.<A_{1}^{p-1},C_{1}^{r-1}>_{tep}+\alpha.f(a_p,c_r).g(t_{a_p},t_{c_r})+ \alpha^2.<A_{1}^{p-1},C_{1}^{r}>_{tep}$.\\

The unique solution is $\alpha=1$ and $\beta=-1$. That is if $<.,.>_{tep}$ is an existing inner product, then necessarily $\alpha=1$ and $\beta=-1$, establishing iv).\\

\textbf{Proof of the converse implication}\\
Let us suppose that i), ii), iii) and iv) are satisfied and show that $<.,.>_{tep}$ is an inner product on $\mathbb{U}^*$.\\

First, by construction, since $f$ and $g$ are symmetric, so is $<.,.>_{tep}$. \\

It is easy to show by induction that $<.,.>_{tep}$ is non-decreasing with the length of its arguments, namely, $\forall A_{1}^p$ and $B_{1}^q$ in $\mathbb{U^*}$,\\
$<A_{1}^{p}, B_{1}^{q}>_{tep}-<A_{1}^{p}, B_{1}^{q-1}>_{tep} \ge 0$. Let $n=p+q$. The proposition is true at rank $n=0$. It is also true if $A_{1}^p=\Omega$, whatever $B_{1}^q$ is, or $B_{1}^q=\Omega$, whatever $<A_{1}^{p}$ is. Suppose it is true at a rank $n \ge 0$, and consider $A_{1}^p \neq \Omega$ and $B_{1}^q \neq \Omega$ such that $p+q=n$.\\
By decomposing $<A_{1}^{p}, B_{1}^{q}>_{tep}$ we get:\\
$<A_{1}^{p}, B_{1}^{q}>_{tep}-<A_{1}^{p}, B_{1}^{q-1}>_{tep} = - <A_{1}^{p-1}, B_{1}^{q-1}>_{tep} + f(a_p,b_q).g(t_{a_p}, t_{b_q}) + <A_{1}^{p-1}, B_{1}^{q}>_{tep}$\\
Since $f(a_p,b_q).g(t_{a_p}, t_{b_q}) > 0$ and the proposition is true by inductive hypothesis at rank $n$, we get that $<A_{1}^{p}, B_{1}^{q}>_{tep}-<A_{1}^{p}, B_{1}^{q-1}>_{tep}) > 0$. By induction the proposition is proved.\\

Let us show by induction on the length of the times series the positive definiteness of $<.,.>_{tep}$.\\
At rank $0$ we have $<\Omega,\Omega>_{tep}=\xi=0$. At rank $1$, let us consider any time series of length $1$, $A_{1}^1$. $<A_{1}^1,A_{1}^1>_{tep}=f(a_1,a_1).g(t_{a_1},t_{a_1})>0$ by hypothesis on $f$ and $g$. Let us suppose that the proposition is true at rank $n>1$ and let consider
 any time series of length $n+1$, $A_{1}^{n+1}$. Then, since $\alpha=1$ and $\beta=-1$,\\
$<A_{1}^{n+1},A_{1}^{n+1}>_{tep}=2.<A_{1}^{n+1},A_{1}^{n}>_{tep} - <A_{1}^{n},A_{1}^{n}>_{tep}+f(a_{n+1},a_{n+1}).g(t_{a_{n+1}},t_{a_{n+1}})$. \\
Since $<A_{1}^{n+1},A_{1}^{n}>_{tep} - <A_{1}^{n},A_{1}^{n}>_{tep} \ge 0$, and $h(A(n+1), A(n+1)>0$, $<A_{1}^{n+1},A_{1}^{n+1}>_{tep}>0$, showing that the proposition is true at rank $n+1$. By induction, the  proposition is proved, which establishes the positive-definiteness of $<.,.>_{tep}$ since $<A_{1}^{p}, A_{1}^{p}>_{tep}=0$ only if $A_{1}^{p}=\Omega$.\\

Let us consider any $\lambda \in \mathbb{R}$, and any $A_{1}^p, B_{1}^q$ in $\mathbb{U^*}$ and show by induction on $n=p+q$ that$<\lambda \otimes A_{1}^p, B_{1}^q>_{tep}=\lambda.<A_{1}^p, B_{1}^q>_{tep}$: \\
The proposition is true at rank $n=0$. Let us suppose that the proposition is true at rank $n \ge 0$, i.e. for all $r\le n$, and consider any pair $A_{1}^p, B_{1}^q$ of time series such that $p+q=n+1$.\\
We have: $<\lambda \otimes A_{1}^p, B_{1}^q>_{tep}=\alpha.<\lambda \otimes A_{1}^p, B_{1}^{q-1}>_{tep}+\beta.<\lambda \otimes A_{1}^{p-1}, B_{1}^{q-1}>_{tep}+f(\lambda\otimes_S a_p, b_q).g(t_{a_p},t_{b_q}) + \alpha.<\lambda \otimes A_{1}^{p-1}, B_{1}^{q}>_{tep}$\\
Since $f$ is linear on $(S, \oplus_S, \otimes_S)$, and since the proposition is true by hypothesis at rank $n$, we get that $<\lambda \otimes A_{1}^p, B_{1}^q>_{tep}=\lambda.\alpha <A_{1}^p, B_{1}^{q-1}>_{tep} + \lambda.\beta.<A_{1}^{p-1}, B_{1}^{q-1}>_{tep}+\lambda.f(a_p, b_q).g(t_{a_p},t_{b_q}) + \lambda.\alpha.<A_{1}^{p-1}, B_{1}^{q}>_{tep}=\lambda.<A_{1}^p, B_{1}^q>_{tep}$.\\
By induction, the proposition is true for any $n$, and we have proved this proposition.

Furthermore, for any $A_{1}^p, B_{1}^q$ and $C_{1}^r$ in $\mathbb{U^*}$, let us show by induction on $n=p+q+r$ that $< A_{1}^p \oplus B_{1}^q, C_{1}^r>_{tep}=< A_{1}^p, C_{1}^r>_{tep}+< B_{1}^q, C_{1}^r>_{tep}$. Let $X_{1}^s$ be equal to $A_{1}^p \oplus B_{1}^q$. The proposition is obviously true at rank $n=0$. Let us suppose that it is true up to rank $n \ge 0$, and consider any $A_{1}^p, B_{1}^q$ and $C_{1}^r$ such that $p+q+r=n+1$. 

Three cases need then to be considered:
\begin{enumerate}[1)]
\item if $X_{1}^{s-1}=A_{1}^{p-1} \oplus B_{1}^{q-1}$, then $t_{a_p}=t_{b_q}=t$ and $< A_{1}^p \oplus B_{1}^q, C_{1}^r>_{tep}=\alpha.< A_{1}^p \oplus B_{1}^q, C_{1}^{r-1}>_{tep} + \beta.< A_{1}^{p-1} \oplus B_{1}^{q-1}, C_{1}^{r-1}>_{tep} +f((a_p+b_q), c_r).g(t,t_{c_r})+ \alpha.< A_{1}^{p-1} \oplus B_{1}^{q-1}, C_{1}^{r}>_{tep}$. Since $f$ is linear on $(S, \oplus_S, \otimes_S)$, and the proposition true at rank $n$, we get the result.
\item if $X_{1}^{s-1}=A_{1}^{p} \oplus B_{1}^{q-1}$, then $t_{a_p} < t_{b_q}=t$ and $< A_{1}^p \oplus B_{1}^q, C_{1}^r>_{tep}=\alpha.< A_{1}^p \oplus B_{1}^q, C_{1}^{r-1}>_{tep} + \beta.< A_{1}^{p} \oplus B_{1}^{q-1}, C_{1}^{r-1}>_{tep} +f(b_q, c_r).g(t,t_{c_r})+ \alpha.< A_{1}^{p} \oplus B_{1}^{q-1}, C_{1}^{r}>_{tep}$. Having $\alpha=1$ and $\beta=-1$ with the proposition supposed to be true at rank $n$ we get the result.
\item if $X_{1}^{s-1}=A_{1}^{p-1} \oplus B_{1}^{q-1}$, we proceed similarly to case 2).
\end{enumerate}

Thus the proposition is true at rank $n+1$, and by induction the proposition is true for all $n$. This establishes the linearity of $<.,.>_{tep}$.\\
This ends the proof of the converse implication and theorem \ref{Main theorem} is therefore established $\square$\\.

The existence of functions $f$ and $g$ entering into the definition of $<.,.>_{tep}$ and satisfying the conditions allowing for the construction of an  inner product on $(\mathbb{U}^*, \oplus, \otimes)$ is ensured by the following proposition:
\begin{prop}
\label{PropExistence}
The functions $f:S^2 \rightarrow \mathbb{R}$ defined as $f(a,b) = <a, b>_S$ where $<.,.>_S$ is an inner product on $(S, \oplus_S, \otimes_S)$ and $g:T^2 \rightarrow \mathbb{R}$ defined as $f(t_a,t_b))=e^{-d(t_a,t_b)}$, where $d$ is a distance defined on $T^2$ and $\nu \in \mathbb{R}^+$, satisfy the conditions required to construct an elastic inner product on $(\mathbb{U^*}, \oplus, \otimes)$.
\end{prop}

The proof of Prop.\ref{PropExistence} is obvious. This proposition establishes the existence of $TEP$ inner products, that we will denote $TEIP$ (Time Elastic Inner Product). Note that $<.,.>_S$ can be chosen to be a $TEIP$ as well, in the case where a second \textit{time elastic} dimension is required. This leads naturally to recursive definitions for $TEP$ and $TEIP$.

\begin{prop}
\label{PropTEPvsEIP}
For any $n \in \mathbb{N}$, and any discrete subset $T=\{t_1, t_2, \cdots, t_n\} \subset \mathbb{R}$, let $\mathbb{U}_{n,\mathbb{R}, T}$ be the set of all time series defined on $\mathbb{R} \times T$ whose lengths are $n$ (the time series in $\mathbb{U}_{n,\mathbb{R}, T}$ are considered to be uniformly sampled). Then, the $TEIP$ on $\mathbb{U}_{n,\mathbb{R}}$ constructed from the functions $f$ and $g$ defined in Prop. \ref{PropExistence} tends towards the Euclidean inner product when $\nu \rightarrow \infty$ if $S$ is an Euclidean space and $<a,b>_S$ is the Euclidean inner product defined on $S$.
\end{prop}

The proof of Prop.\ref{PropTEPvsEIP} is straightforward and is omitted. Prop.\ref{PropTEPvsEIP} shows that a $TEIP$ generalizes the classical Euclidean inner product.\\

\section{Some applications}

We present in the following sections some applications to highlight the properties of Time Elastic Vector Spaces ($TEVS$).

\subsection{Distance in $TEVS$}

The following proposition provides $\mathbb{U}^*$ with a norm and a distance, both induced by a $TEIP$.

\begin{prop}
\label{TEP_distance}
For all $A_{1}^{p} \in \mathbb{U}^*$, and any $<.,.> TEIP$  defined on $(\mathbb{U}^*, \oplus, \otimes)$  $\sqrt{<A_{1}^{p},A_{1}^{p}>}$ is a norm on $\mathbb{U}^*$.\\
For all pair $(A_{1}^{p},B_{1}^{q}) \in (\mathbb{U}^*)^2$, and any $TEIP$  defined on $(\mathbb{U}^*, \oplus, \otimes)$,  $\delta(A_{1}^{p},B_{1}^{q})= \sqrt{<A_{1}^{p} \oplus (-1.\otimes B_{1}^{q}), A_{1}^{p} \oplus (-1.\otimes B_{1}^{q})>}$ defines a distance metric on $\mathbb{U}^*$.
\end{prop}

The proof of Prop. \ref{TEP_distance} is straightforward and is omitted. \\

\subsection{Orthogonalization in $TEVS$}

To exemplify the effect of elasticity in $TEVS$, we give below the result of the Gram-Schmidt orthogonalization algorithm for two families of independent time series. The first family is composed of uniformly sampled time series having increasing lengths. The second family (a sine-cosine basis) is composed of uniformly sampled time series, all of which have the 
same length.  \\

The tests which are described in the next sections were performed on a set $\mathbb{U}^*$ of discrete time series whose elements are defined on $(\mathbb{R}-\{0\} \times [0;1])^2$ using the following $TEIP$:\\

\begin{eqnarray}
\label{Eq_teip}
\begin{array}{ll}
 &<A_{1}^p, B_{1}^q>_{teip} =  \\
 & \sum \left\{
   \begin{array}{ll}
     <A_{1}^{p},B_{1}^{q-1}>_{teip}  \\
     - <A_{1}^{p-1},B_{1}^{q-1}>_{teip} + a(p)b(q)\cdot e^{-\nu.|t_{a_p}-t_{b_q}|} \\
     <A_{1}^{p-1},B_{1}^{q}>_{teip}  \\
   \end{array}
   \right.
  \end{array}
\end{eqnarray}

\subsubsection{Orthogonalization of an independent family of time series with increasing lengths}
The family of time series we are considering is composed of $11$ time series uniformly sampled, whose lengths are $11$ samples: 
\begin{eqnarray}
\label{BasisOfIncreasingLength}
\begin{array}{ll}
 (1,0)\\
 (\epsilon,0)(1,1/10)\\
 (\epsilon,0)(\epsilon,0)(1,1/10)\\
 \cdots\\
 (\epsilon,0)(\epsilon,1/10)(\epsilon,2/10) \cdots (1,1)\\
\end{array}
\end{eqnarray}

\begin{figure*}[]
\centering
\includegraphics[scale=0.35]{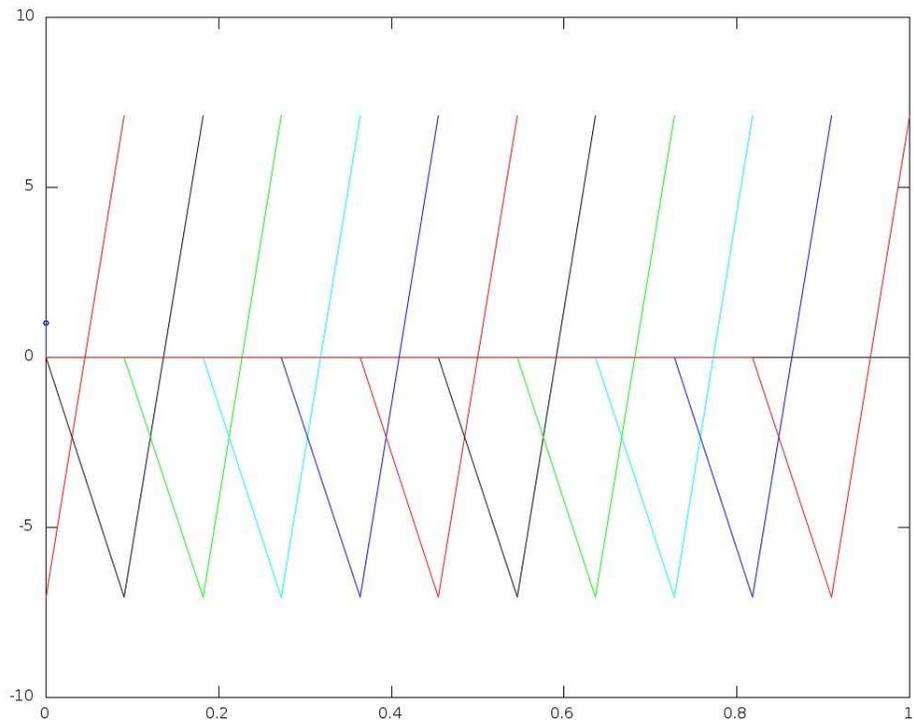}
\caption{Result of the orthogonalization of the family of length time series defined in Eq.\ref{BasisOfIncreasingLength} using $\nu=.01$: except for the first \textit{spike} located at \textit{time} $0$, each original \textit{spike} is replaced by two \textit{spikes}, one negative the other positive.} 
\label{fig:spk_orthogonal}
\end{figure*}

Since, the zero value cannot be used for the space dimension, we replaced it by $\epsilon$, which is the smallest non zero positive real for our test machine (i.e. $2^{-1074}$). The result of the Gram-Schmidt orthogonalization process using $\nu=.01$ on this basis is given in Fig.\ref{fig:spk_orthogonal}.

\subsubsection{Orthogonalization of a sine-cosine basis}

\begin{figure*}[!h]
\centering
\includegraphics[scale=0.65]{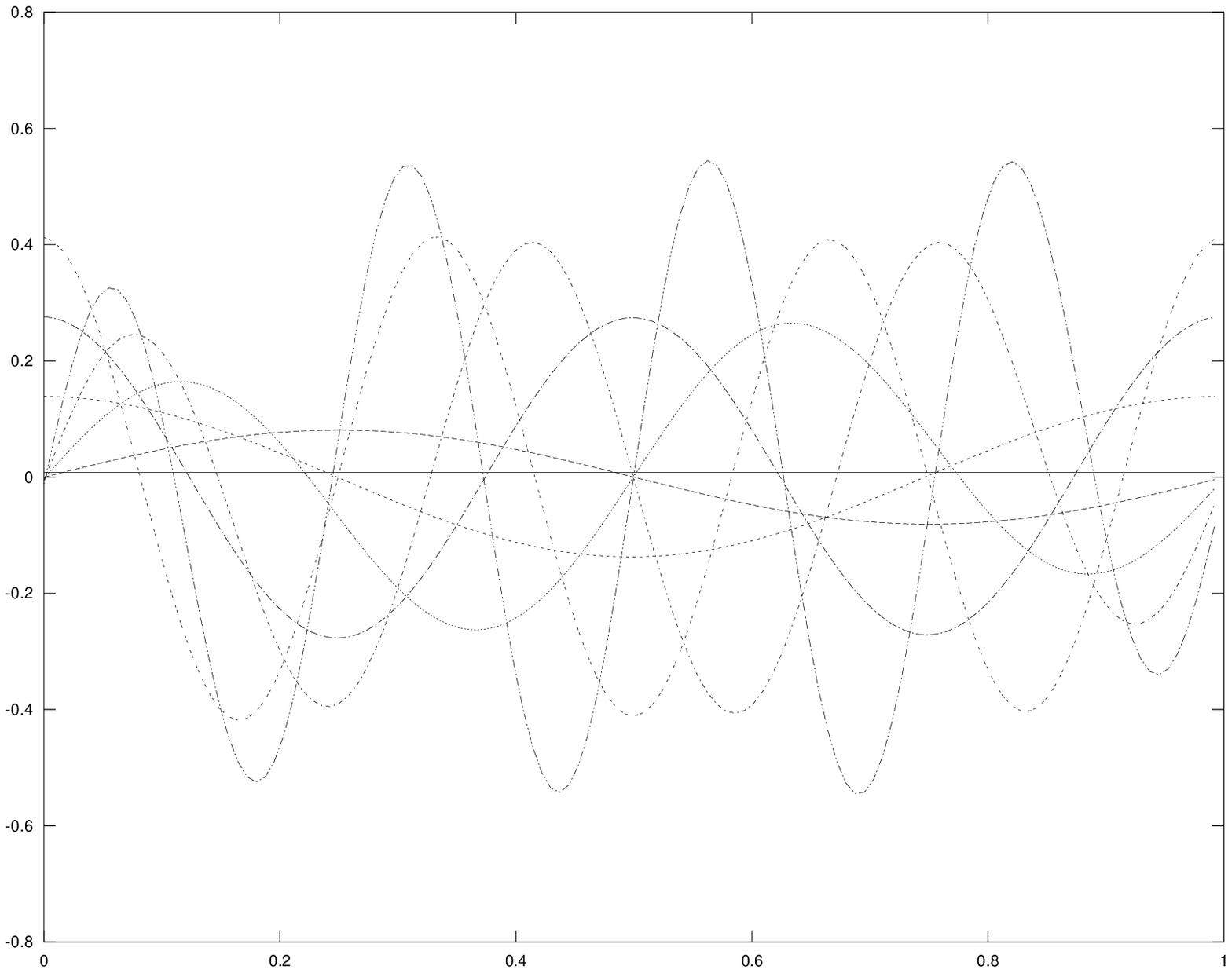}
\caption{Orthogonalization of the \textit{sine-cosine} basis using $\nu=.01$: the waves are slightly deformed jointly in amplitude and in frequency. For readability of the figure, we have presented the 8 first components} 
\label{fig:sin_orthogonal}
\end{figure*}
An orthonormal family of discrete sine-cosine functions is not anymore orthogonal in a $TEVS$.
The result of the Gram-Schmidt orthogonalization process using $\nu=.01$ when applied on a discrete sine-cosine basis is given in Fig.\ref{fig:sin_orthogonal}, in which only the 8 first components are displayed. The lengths of the waves are 128 samples.

\subsection{Kernel methods in $TEVS$}
A wide range of literature exists on kernels, among which \cite{TWED:BergChristensenRessel84}, \cite{TWED:Scholkopf01} and \cite{TWED:Shawe04} present some large syntheses of  major results. 
\begin{defn}
\label{Kernels}
A kernel on a non empty set $U$ refers to a complex (or real) valued symmetric function $\varphi(x,y) : U \times U \rightarrow \mathbb{C}$ (or $\mathbb{R}$).  
\end{defn}

\begin{defn}
\label{Definite Kernels}
Let $U$ be a non empty set. A function $\varphi: U \times U \rightarrow \mathbb{C}$ is called a positive (resp. negative) definite kernel if and only if it is Hermitian 
(i.e. $\varphi(x,y)=\overline{\varphi(y,x)}$ where the \textit{overline} stands for the conjugate number) for all $x$ and $y$ in $U$ and $\sum_{i,j=1}^n c_i \bar{c_j} \varphi(x_i, x_j) \ge 0$ (resp. $\sum_{i,j=1}^n c_i \bar{c_j} \varphi(x_i, x_j) \le 0$), 
for all $n$ in $\mathbb{N}$, $(x_1,x_2, ..., x_n) \in U^n$ and $(c_1,c_2,...,c_n) \in \mathbb{C}^n$.\\
\end{defn}

\begin{defn}
 \label{Conditionally Definite Kernels}

Let $U$ be a non empty set. A function $\varphi: U \times U \rightarrow \mathbb{C}$ is called a conditionally positive (resp. conditionally negative) definite kernel if and only if it is Hermitian 
(i.e. $\varphi(x,y)=\overline{\varphi(y,x)}$ for all $x$ and $y$ in $U$) and $\sum_{i,j=1}^n c_i \bar{c_j} \varphi(x_i, x_j) \ge 0$ (resp. $\sum_{i,j=1}^n c_i \bar{c_j} \varphi(x_i, x_j) \le 0$),
for all $n \ge 2$ in $\mathbb{N}$, $(x_1,x_2, ..., x_n) \in U^n$ and $(c_1,c_2,...,c_n) \in \mathbb{C}^n$ with $\sum_{i=1}^n c_i=0$. \\
\end{defn}

In the last two above definitions, it is easy to show that it is sufficient to consider mutually different elements in $U$, i.e. collections of distinct elements $x_1, x_2, ..., x_n$. \\

\begin{defn}
 \label{matrix and kernel}
A positive (resp. negative) definite kernel defined on a finite set $U$ is also called a positive (resp. negative) semidefinite matrix. 
Similarly, a positive (resp. negative) conditionally definite kernel defined on a finite set is also called a positive (resp. negative) conditionally semidefinite matrix.
\end{defn}

\subsubsection{Definiteness of $TEIP$ based kernel}

\begin{prop}
\label{TEIPKernelDefiniteness}
A $TEIP$ is a positive definite kernel.\\
\end{prop}

The proof of Prop. \ref{TEIPKernelDefiniteness} is straightforward and is omitted. \\

\subsubsection{SVM classification using a $TEP$ based kernel}
 In \cite{TWED:Marteau2010}, $<.,.>_{twip_2}$ (Eq.\ref{eqTWIP2}) have been experimented on a classification task using a SVM classifier on 20 datasets containing times series uniformly sampled and having the same lengths inside each dataset. On the same data, we get similar results for $<.,.>_{teip}$ (Eq.\ref{Eq_teip}) and do not report them in this paper. The benefit of introducing some time elasticity, controlled using the parameter $\nu$ is quite clear when comparing the classification error rates obtained using a Gaussian kernel exploiting the distance derived from $<.,.>_{teip}$ (Prop. \ref{TEP_distance}) with the classification error rates obtained using a Gaussian kernel exploiting the Euclidean distance.

\subsection{Elastic Cosine similarity in $TEVS$, with application to symbolic (e.g. textual) information retrieval}
Similarly to the definition of the cosine of two vectors in Euclidean space, we define the elastic cosine of two sequences by using any $TEP$ that satisfies the conditions of theorem \ref{Main theorem}.

\begin{defn}
 \label{elastic cosine}
Given two sequences, $A$ and $B$, the elastic cosine similarity of these two sequences is given using a time elastic inner product $<X,Y>_{e}$ and the induced norm $\|X\|_e=\sqrt{<X,X>_e}$ as\\
    $\textit{similarity} = \cos_e(\theta) = {<A \cdot B>_{e} \over \|A\|_e \|B\|_e}$
\end{defn}

In the case of textual information retrieval, namely text matching, the timestamps variable coincides with the index of words into the text, and the spatial dimensions encode the words into a given dictionary. For instance, each word can be represented using a vector whose dimension is the size of the set of concepts (or senses) that cover the conceptual model associated to the dictionary and each coordinate selected into $[0;1]$ encodes the degree of presence of the concept or senses into the considered word. In that case, the elastic cosine similarity measure takes value into $[0;1]$, $0$ indicating the lowest possible similarity value between two texts and $1$ the greatest possible similarity value between two texts. The elastic cosine similarity takes into account the order of occurrence of the words into a text which could be an advantage compared to the Euclidean cosine measure that does not cope with the words ordering.

Let us consider the following elastic inner product dedicated to text matching. In the following definition, $A_{1}^{p}$ and $B_{1}^{q}$ are sequences of words that represent textual content.
\begin{defn}
\label{eqEIP_TM}
\begin{eqnarray}
\begin{array}{ll}
<A_{1}^{p}, B_{1}^{q}>_{teip_{tm}}=   \\
\hspace*{2mm} \sum \left\{ 
		\begin{array}{ll}
     <A_{1}^{p-1},B_{1}^{q}>_{teip_{tm}} \\
     -<A_{1}^{p-1},B_{1}^{q-1}>_{teip_{tm}} + e^{-\nu.|t_{a(p)}-t_{b(q)}|}\delta(a(p), b(q))\\
     <A_{1}^{p},B_{1}^{q-1}>_{teip_{tm}} \\
   \end{array}
   \right.
  \end{array}
\end{eqnarray}
where $\delta(x,y)=1$ if $x=y$ ($x$ and $y$ identify the same word), $0$ otherwise, and $\nu$ a \textit{time stiffness} parameter.\\
\end{defn}

\begin{prop}
 \label{DTEIP_vs_vectorModel}
For $\nu=0$, the elastic inner product defined in Eq.\ref{eqEIP_TM} coincides with the euclidean inner product between two vectors whose coordinates correspond to term frequencies observed into the $A_{1}^{p}$ and $B_{1}^{q}$ text sequences. If, we change the definition of $\delta$ by the $\delta(x,y)=IDF(x)$ if $x=y$, $0$ otherwise, where $IDF(x)$ is the inverse document frequency of term $x$ into the considered collection, then for $\nu=0$, $<A_{1}^{p}, B_{1}^{q}>_{teip_{tm}}$ coincides with the euclidean inner product between two vectors whose coordinates correspond to the TF-IDF (term frequency times the inverse document frequency) of terms occurring into the $A_{1}^{p}$ and $B_{1}^{q}$ text sequences.
\end{prop}

The proof of proposition \ref{DTEIP_vs_vectorModel} is straightforward an is omitted.\\

Thus, the elastic cosine measure derived from the elastic inner product defined by Eq.\ref{eqEIP_TM} generalizes somehow the cosine measure implemented in the vector model \cite{TWED:SaltonG84} and commonly used in the text information retrieval community.
 
\section{Conclusion}

This paper proposed what we call a family of \textit{time elastic inner products} able to cope with non-uniformly sampled time series of various lengths, as far as they do not contain the \textit{zero} value. These constructions allow one to embed any such time series in a single vector space, that some how generalizes the notion of Euclidean vector space. The recursive structure of the construction offers the possibility to manage several \textit{time elastic} dimensions. Some applicative benefits could be expected in time series analysis when time \textit{elasticity} is an issue, for instance in the field of numeric or symbolic sequence data mining.




\bibliography{EVS}{}
\bibliographystyle{model1a-num-names}







\end{document}